\documentclass{aa}
\usepackage{graphicx}
\begin{document}
   \title{\textbf{Soft X-ray coronal spectra at low activity levels observed by RESIK}}
   \titlerunning{RESIK low-activity spectra}
   \author{B. Sylwester,
          \inst{1}
          J. Sylwester
          \inst{1}
          \and
          K.J.H. Phillips
          \inst{2}}
   \offprints{B. Sylwester}
 %  \authorrunning{B. sylwester}
   \institute{Space Research Center, Polish Academy of Sciences,
                Kopernika 11, 51-622  Wroc{\l}aw, Poland\\
              \email{(bs, js)@cbk.pan.wroc.pl
              } \and
                UCL-Mullard Space Science Laboratory,
                Holmbury St. Mary, Dorking, Surrey RH5 6NT, UK\\
              \email{kjhp@mssl.ucl.ac.uk}
                }
   \date{Received ; accepted }

% \abstract{}{}{}{}{}
% 5 {} token are mandatory

  \abstract
{The quiet-Sun X-ray emission is important for deducing coronal
heating mechanisms, but it has not been studied in detail since the
emph{Orbiting Solar Observatory} (\emph{OSO}) spacecraft era. Bragg
crystal spectrometer X-ray observations have generally concentrated
on flares and active regions. The high sensitivity of the RESIK
(REntgenovsky Spectrometer s Izognutymi Kristalami) instrument on
the \emph{CORONAS-F} solar mission has enabled the X-ray emission
from the quiet corona to be studied in a systematic way for the
first time. }
{Our aim is to deduce the physical conditions of the non-flaring
corona from RESIK line intensities in several spectral ranges using
both isothermal and  multithermal assumptions.}
{We selected and analyzed spectra in 312 quiet-Sun intervals in
January and February 2003, sorting them into 5 groups according to
activity level. For each group, the fluxes in selected spectral
bands have been used to calculate values parameters for the best-fit
that lead to a intensities characteristic of each group. We used
both isothermal and multitemperature assumptions, the latter
described by differential emission measure (DEM) distributions.
RESIK spectra cover the wavelength range ($3.3-6.1$~\AA). This
includes emission lines of highly ionized  Si, S, Cl, Ar, and K,
which are suitable for evaluating temperature and emission measure,
were used.   }
{The RESIK spectra during these intervals of very low solar activity
for the first time provide information on the temperature structure
of the quiet corona. Although most of the emission seems to arise
from plasma with a temperature between 2~MK and 3~MK, there is also
evidence of a hotter plasma ($T \sim 10$~MK) with an emission
measure 3 orders smaller than the cooler component. Neither coronal
nor photospheric element abundances appear to describe the observed
spectra satisfactorily.}
   {}


 \maketitle
%
%________________________________________________________________

\section{Introduction}
The RESIK X-ray spectrometer on the Russian {\it CORONAS-F} solar
orbiting mission (circular polar orbit: altitude 550~km, period 96
minutes) obtained numerous high-resolution flare and active-region
spectra in the $3.3-6.1$~\AA\ range over the period
August~2001~--~May~2003. The RESIK instrument (Sylwester et al.,
2005) was a bent crystal spectrometer with four spectral channels in
which solar X-ray emission was diffracted by crystal wafers made of
silicon (Si 111, $2d = 6.27$~\AA) and quartz (Qu $10\bar 10$, $2d =
8.51$~\AA). Although most previous spacecraft crystal spectrometers
suffered from the strong instrumental backgrounds caused by
fluorescence of the crystal material, RESIK had a system of
pulse-height analyzers enabling primary solar photons to be
distinguished from secondary photons produced by fluorescence. The
background could thus be practically eliminated for much of the
period 2003 January~--~March when the non flaring solar X-ray
activity was often below C1 class as measured by the {\it GOES}
1~--~8~\AA\ sensor. The sensitivity of RESIK was maximized by not
having a collimator placed in front of the crystals. This, like the
equivalent Bragg Crystal Spectrometer on the {\it Yohkoh} spacecraft
(operational 1991--2001), produced some spectral confusion when two
simultaneous bright sources were present on the Sun, but in practice
this rarely occurred. The low-activity corona gives rise to X-ray
line profiles in RESIK spectra having large line widths through
spatial broadening.

Several analyses have been done on spectra of solar flares from
RESIK (Sylwester et al., 2006; Chifor et al., 2007; Sylwester,
Sylwester and Phillips, 2008), but here we report on spectra
obtained during a period of sustained low solar activity, which have
high statistical quality because of the relatively high sensitivity
of RESIK. The observed spectral shapes including continua are
available for analysis of the temperature structure of the emitting
regions, specifically the differential emission measure as a
function of electron temperature $T$; and from this, the absolute
abundances of the elements giving rise to the spectral lines can be
assessed.
%%%%%%%%%%%%%%%%%%%%%%%% Begin of Figure 1  TWO COLUMN   %%%%%%%%%%%%%%%%%%%%%%
   \begin{figure}[t]
    \vspace{3mm}
    \hspace{10mm}
  \centering
   \includegraphics[width=5.5cm]{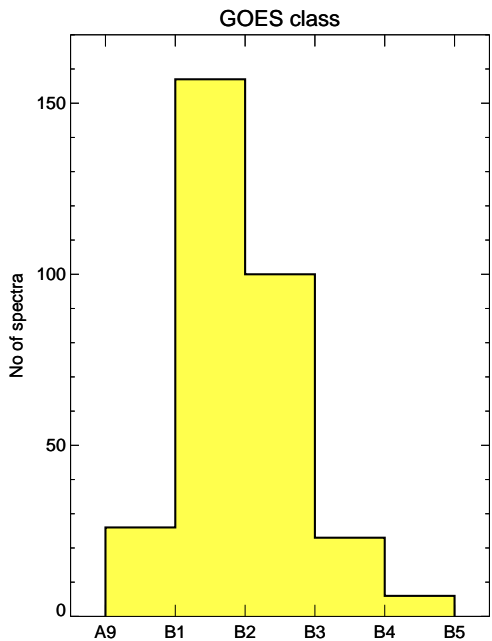}
   \vspace{3mm}
   \caption{Histogram showing the respective \emph{GOES} activity for selected  312 time intervals (January-February 2003).}
              \label{FigGam}
    \end{figure}
%%%%%%%%%%%%%%%%%%%%%%% end of Figure 1 %%%%%%%%%%%%%%%%%%%%%%

Quiet-Sun X-ray spectra have not received nearly as much attention
in the past as those from flares, but some properties of the quiet
Sun have been widely studied using its ultraviolet emission, which
has been measured by the following experiments: Orbiting Solar
Observatory \emph{OSO} series (1962-1975; see for instance Dupree
and Reeves, 1971; Dupree et al., 1973), \emph{Aerobee} rocket
spectrometer (1969, see Malinovsky and Heroux, 1973), 9 months of
the \emph{Skylab} mission (May 1973-February 1974, see Vernazza and
Reeves, 1978), the series of 9 Solar EUV Rocket Telescope and
Spectrograph (SERTS) flights (the first in 1989, see Brosius et al.,
1996, 1998), the \emph{SOHO} mission (starting in December 1995, CDS
and SUMER data, see Warren, Mariska and Lean, 1998). The data
obtained have been used as well to construct a reference atlas of
quiet-Sun ultraviolet radiation (Curdt, Landi and Feldman, 2004)
from which  differential emission measure (DEM) distributions can be
inferred. Brosius et al. (1996) calculated DEM distributions for
quiet-Sun conditions during  two observing periods near the maximum
of Cycle~21 (1991 May 7 and 1993 August 17) based on SERTS data.
Their DEM solutions included a power law in the temperature range
$6.3\times 10^4$~K--$5.0\times 10^5$~K with hot plasma as shown by a
localized maximum at 5~MK. A DEM with similar distribution was
obtained by Kretzschmar, Lilensten, and Aboudarham (2004) based on
\emph{SOHO} SUMER data, having a power-law shape in the range
$2.0\times 10^4$~K--$2.0\times 10^5$~K with a high-temperature bump
at 1.1~MK. Ralchenko, Feldman, and Doschek (2007) have studied
quiet-corona spectra as observed by SUMER during the 2000 June
13--19 period, deducing that the observed line intensities can be
satisfactorily described by a model with two Maxwellian electron
distributions, a first population corresponding to an isothermal
temperature of $\sim 1.3\times 10^6$~K, and a second, smaller
population ($\sim 5\%$) of hot ({300--400}~eV) electrons accounting
for the intensities of highly charged Ar and Ca ion lines observed
by SUMER.
%%%%%%%%%%%%%% begin Fig. 2 %%%%%%%%%%%%%%%%%%%%%%%%%%%%%%%%%%%%%%%%%
   \begin{figure}[t]
   \centering
   \includegraphics[width=8cm]{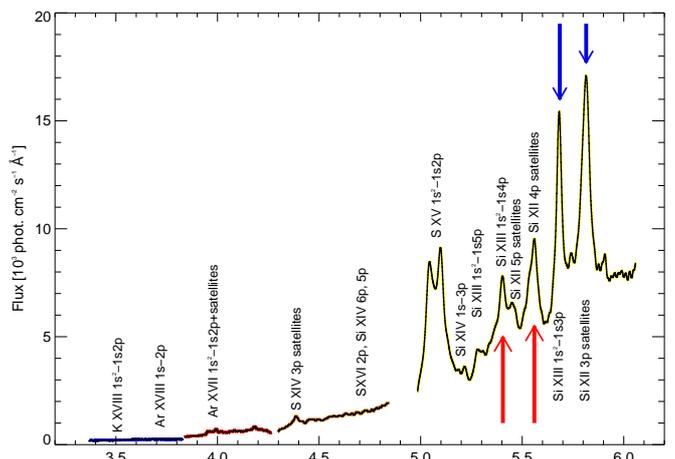}
   \vspace{-0mm}
      \caption{
      Composite of four full-disk EIT/SOHO images obtained in the wavelength bands
       304~\AA, $171$~\AA, $195$~\AA, and $284$~\AA, on 2003~February~24 around
       19:00~UT. This was the time of lowest activity in the period of the RESIK quiet-Sun spectra analyzed in this work.
      }
         \label{FigVibStab}
   \end{figure}
%%%%%%%%%%%%%%%%%%%%%%% end Fig. 2 %%%%%%%%%%%%%%%%%%%%%%%%%%%%%%%%
The DEM analysis was also performed using \emph{SOHO} Coronal
Diagnostic Spectrometer (CDS) data for the internetwork, network,
and bright network regions of the quiet Sun by O'Shea et al.~(2000).
They find that the DEM distributions differ in each region over the
temperature range 0.25~--~1~MK.

Recently, Young et al. (2007) have published a quiet-Sun extreme
ultraviolet (EUV) spectrum in the ranges {170--211}~\AA\ and
{246--292}~\AA, as observed on 2006 December 23 by the \emph{Hinode}
EUV Imaging Spectrometer (EIS).
% Table 1
\begin{table}[h]
\caption{Wavelength intervals used for DEM studies}  % title of Table
\label{table:1}      % is used to refer this table in the text
\centering                          % used for centering table
\begin{tabular}{c c c}        % centered columns (4 columns)
\hline\hline                 % inserts double horizontal lines
No & Range [\AA] & Dominant contributor  \\    % table heading
\hline                        % inserts single horizontal line
\vspace{-3mm}
     &               &                                                       \\
    1 & 3.40~-~3.62 & cont. + K~{\sc xviii }$2p$+sat.                      \\
    2 & 3.62~-~3.80 & cont. + Ar~{\sc xviii }$2p$, S~{\sc xvi }$4p, 5p$    \\
    3 & 3.85~-~4.10 & cont. + Ar~{\sc xvii }$2p$+sat, S~{\sc xv }$4p$      \\
    4 & 4.10~-~4.25 & cont. + S~{\sc xv }sat.                              \\
    5 & 4.35~-~4.43 & cont. + S~{\sc xv }$3p$+sat                          \\
    6 & 4.43~-~4.52 & cont. + Cl~{\sc xvi }$2p$+sat                        \\
    7 & 4.68~-~4.74 & cont. + S~{\sc xvi }$2p$, + Si~{\sc xiv }$8p$        \\
    8 & 4.74~-~4.81 & cont. + Si~{\sc xiv }$6p$                            \\
    9 & 5.00~-~5.13 & S~{\sc xv }$2p$+sat + cont.                          \\
   10 & 5.26~-~5.35 & Si~{\sc xiii }$5p$+sat.  + cont.                     \\
   11 & 5.36~-~5.50 & Si~{\sc xiii }$4p$+sat.  + cont.                     \\
   12 & 5.50~-~5.63 & Si~{\sc xii } sat. + cont.                           \\
   13 & 5.64~-~5.72 & Si~{\sc xiii }$3p$ + cont.                           \\
   14 & 5.73~-~5.86 & Si~{\sc xii } sat. + cont.                           \\
   15 & 5.90~-~6.00 &  continuum                                           \\
\hline                                   %inserts single line
\end{tabular}
\vspace{-5mm}
\end{table}

The quiet-Sun hard X-ray emission observed by \emph{RHESSI} has been
examined by Hannah et al. (2007) using a fan-beam modulation
technique during seven periods of off-pointing of the \emph{RHESSI}
spacecraft between 2005 June and 2006 October. They established new
upper limits on the 3--200~keV X-ray emission for when the
\emph{GOES} level of activity was below A1 class, updating much
earlier measurements (Peterson et al., 1966). Schmelz et al. (2009)
have investigated the emission of a nonflaring active region based
on the ten filters of the \emph{Hinode} X-ray Telescope data using
two independent algorithms to reconstruct the differential emission
measure distribution. In addition to the typical low-temperature
emission measure ($T<5$~MK), they find a very hot component ($\sim
30$~MK) with small emission measure. These findings have recently
been modified by Schmelz 2009, priv. comm.: the temperature of the
hotter component is now much lower, around 10~MK. Reale et al.
(2009) have investigated the \emph{Hinode} XRT data averaged over
one hour during a nonflaring period (2006 November~12), finding a
hotter component (temperature $\sim 6.3$~MK) corresponding to a
nonflaring active region. These observational results are supported
by the theoretical work of Klimchuk et al. (2008) who use
hydrodynamic simulations of nanoflares to predict a small amount of
hot plasma in addition to the dominant 2--3~MK plasma component.

The RESIK spectra recorded during solar minimum can bridge the gap
between the results obtained from ultraviolet and X-ray images and
spectra and the models of coronal plasma heating. In earlier work,
we analyzed RESIK spectra to determine the conditions of flaring
plasmas, but here we apply the same techniques of analyzing emission
during low-level periods to deduce the properties of the quiet-Sun
corona.
\section{RESIK spectra selection and isothermal analysis}
%%%%%%%%%%%%%% begin Fig. 3  ONE COLUMN  %%%%%%%%%%%%%%%%%%%%%%%%%%%%%%%%%%%%%%%%%
   \begin{figure*}[t]
   \centering
   \includegraphics[width=15.5 cm]{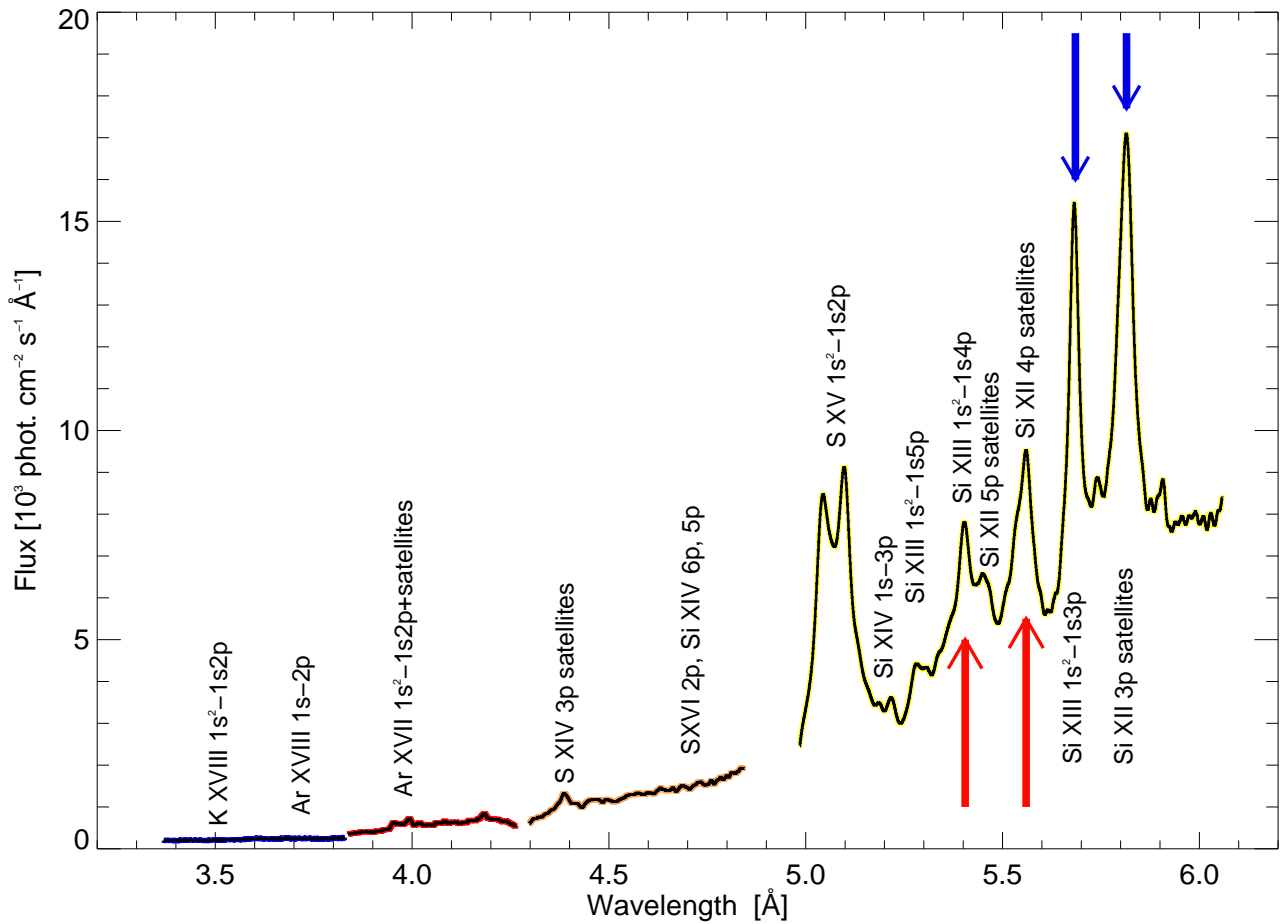}
   \vspace{9mm}
      \caption{The absolute RESIK spectrum averaged over 312 quiet Sun intervals selected between 2003 January and March.
       The most prominent lines are identified.}
         \label{FigVibStab}
   \end{figure*}
%%%%%%%%%%%%%%%%%%%%%%% end Fig. 3 %%%%%%%%%%%%%%%%%%%%%%%%%%%%%%%%
We selected 312 time intervals, each 5~min in duration, for which
solar activity levels in the first quarter of 2003 were at their
lowest. The measurements were taken outside intervals of increased
background owing to passages of the {\it CORONAS-F} spacecraft
through polar ovals or the South Atlantic Anomaly. Early in 2003,
the RESIK instrument characteristics (high voltage and amplitude
discrimination) were set at their optimum values. Each spectrum
analyzed was integrated over a 302~s interval, the longest
selectable period for a low-flux operating mode, so the  overall
data accumulation time was 26.2 hours. The spectra correspond to
\emph{GOES}  activity levels from $\sim$A9 to $\sim$B5. We found it
useful to group the observed spectra according to the activity level
at the time that they were recorded, so that the summed spectra for
any group has an improved statistical quality. The division into
groups was chosen according to activity level steps of 0.1 in the
logarithm (0.1~dex).

In Fig.~1 we show the distribution of spectra with \emph{GOES}
classes. The lowest class includes spectra from the \emph{GOES}
class between A9 and B1 (1--8~\AA\ flux from $9\times 10^{-8}$ to
$1\times 10^{-7}$~W\,m$^{-2}$), attained between 2003~February~23
and February~25. Although flaring periods were always carefully
avoided in selecting spectra, it was impossible to avoid the
presence of active regions, as RESIK operated close to the maximum
of Cycle~23.

A typical pattern of activity during the time of the selected
spectra collection is illustrated (Fig.~2) with a set of full-disk
EUV images from the Extreme-ultraviolet Imaging Telescope (EIT;
Delaboudiniere et al., 1995) aboard the Solar and Heliospheric
Observatory (\emph{SOHO}) on 2003 February 24 around 19:00~UT. The
images are in the spectral bands $304$~\AA\ (He~{\sc ii}),
$171$~\AA\ (Fe~{\sc ix-x}), $195$~\AA\ (Fe~{\sc xii}), and
$284$~\AA\ (Fe~{\sc xv}), which are sensitive to temperatures of
about 80000~K, 1.3~MK, 1.6~MK, and 2~MK, respectively. Weak
active-region emission is visible in all passbands. This is also
true  for all other times when RESIK spectra were selected, so our
results therefore apply to times when the corona was quiet and also
when there were weak active regions. In Fig.~3, the average of all
the 312 spectra is shown, with spectral lines or their expected
wavelengths identified. The background is a true solar continuum.
Identification of main contributing lines is provided in Table~1.

For these quiet-Sun spectra, there is practically no line emission
in channel 1 (shown in blue), though for flare spectra, the
prominent triplet of lines due to K~{\sc xviii} (3.53--3.57~\AA) is
always present (Sylwester et al., 2006). Similarly, the Ar~{\sc
xvii} (3.95--4.00~\AA) triplet in channel 2 (dark red) is barely
distinguished but the lines are very prominent during flares. Only
weak line emission is visible in channel 3 (orange), identifiable
with  dielectronic satellites to the parent $1s-3p$ (\emph{w3})
transition line of the H-like S~{\sc xvi} ion. However, channel 4
(yellow) is rich in strong lines formed by relatively cool plasma
($T < \sim5$~MK).

The He-like S~{\sc xv} line triplet (5.05-5.10~\AA) is prominent but
with only two components resolved owing to increased line widths.
More prominent line features correspond to Si~{\sc xiii} transitions
$1s^2 - 1s4p$ and $1s^2 - 1s3p$ ($w4$, $w3$) at 5.40 and 5.68~\AA,
with nearby Si~{\sc xii} dielectronic satellites at 5.56 and
5.82~\AA\ ($d4$, $D$). The $w4$, $d4$ lines are marked by red
arrows, the $w3$, $D$ lines by blue arrows. Ratios of the flux in an
Si~{\sc xii} dielectronic satellite feature to that in the parent
Si~{\sc xiii} resonance line is very sensitive to temperature in the
range 2--5~MK (Phillips et al., 2006). Here we used the ratio $D/w3$
to determine the cooler plasma component temperature characterizing
the general coronal emission that the selected RESIK spectra are
sensitive to. To do this, we fitted Gaussian profiles to the lines
(continuum level subtracted) and the flux ratios compared with
theoretical values. The derived ``isothermal'' temperatures ($T_{\rm
D/w3}$ raw) are given in Table~2  for each activity class. The
temperatures range from 3.1~MK to $\sim 3.8$~MK for the lowest and
highest activity classes. They are much lower than those
corresponding to flare temperatures (Phillips et al., 2006), even
those late in the decay phase. It is striking to see that the
intensity of the satellite line $D$ is much higher than the parent
line $w3$, even for a medium-$Z$ ($Z=14$) element like Si.
Generally, this has been seen in flare spectra only for Fe~{\sc
xxiv} satellites near Fe~{\sc xxv} lines (near 1.9~\AA: $Z$ for Fe =
26): note that the flux ratio scales as $Z^4$.

A simple way of deducing the physical characteristics of an emitting
plasma is by the ratio of fluxes in broad-band filters. For RESIK,
the corresponding technique is to take the flux ratio of the total
amount of emission in channels 3 and 4. An isothermal assumption is
often too crude to be useful (as will be shown later), and physical
interpretation of derived values can be ambiguous (Sylwester, 1990).
The results of the analysis are given in Table~2 (denoted by ``3/4''
entries). The $T_{3/4}$ values clearly decrease with increasing
activity levels, against expectation, whereas the values of
$T_{D/w3}$ increase. This can be understood in terms of the DEM
analysis presented later. The most commonly used diagnostics of
physical conditions in the coronal thermal plasma component rely on
interpreting the flux ratios measured in the two standard
\emph{GOES} bands. This standard approach (available, e.g., in the
\emph{GOES} section of the IDL \emph{SolarSoft} package: Freeland \&
Handy 1998) was used here, and values of temperatures and emission
measures are given in Table~2 as averages over corresponding
activity groups (denoted by subscript G). It should be stressed,
however, that the \emph{GOES} values derived for the lower intensity
classes may be somewhat biased by problems intrinsic to measurements
of very low flux levels. The results shown in Table~2 indicate that
$T_{\rm G}$ decreases with activity class, as with the values of
$T_{3/4}$. With average values of $T$ and $EM$, we may estimate the
total thermal energy  in the soft X-ray emitting component. It has
been shown by Sylwester et al. (2008) that this energy content can
be expressed as

\begin{equation}
E_{th} = 3k T\, N_e\, V = \sqrt{V}\, ThM~~~~~~   {\rm ergs,}
\end{equation}
\noindent where $V$ is the emitting volume and $ThM$, a
``thermodynamic measure'', is defined by

\begin{equation}
ThM = 3k T \,\sqrt{EM}~~~~~~~~~~~~{\rm g~cm}^{1/2}\,{\rm s}^{-2}.
\end{equation}
The use of $ThM$ is convenient as its variation directly reflects
the changes of thermal energy, provided the emitting volume does not
change substantially. Derived values of $ThM$ are given in Table~2.
It is seen that, in spite of significant differences between the
values of $T_{3/4}$ and $T_{G}$ as well $EM_{3/4}$ and $EM_{G}$,
values of $ThM_{3/4}$ and $ThM_{G}$ agree quite well. This is an
indirect illustration that the value of $ThM$ is relatively
independent of the uncertainties in determinations of $T$ and $EM$
(the uncertainties cancel in the product).
%%%%%%%%%%%%%% begin Fig. 4  Two COLUMNS  %%%%%%%%%%%%%%%%%%%%%%%%%%%%%%%%%%%%%%%%%
   \begin{figure*}
   \centering
   \includegraphics[width=16.5cm]{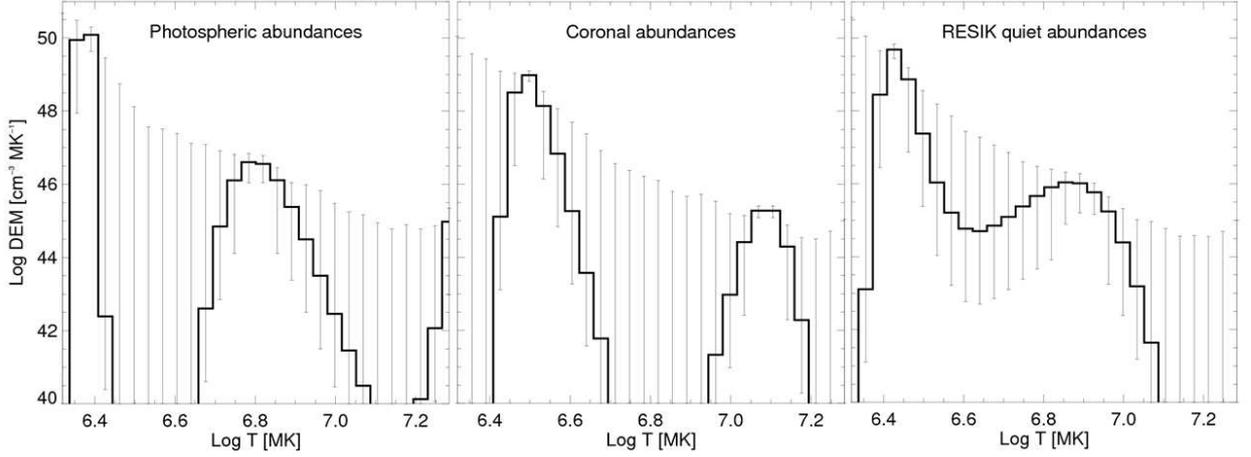}
   \vspace{5mm}
      \caption{The differential emission measure distribution calculated from the fluxes in 15 passbands
      of the averaged nonflaring RESIK spectrum. The results are shown for assuming photospheric (left) coronal (middle) and optimized for the observed spectrum (right) quiet elemental abundances.
}
         \label{FigVibStab}
   \end{figure*}
%%%%%%%%%%%%%%%%%%%%%%% end Fig. 4 %%%%%%%%%%%%%%%%%%%%%%%%%%%%%%%%

\section{Differential emission measure determinations}

It is generally known that the isothermal approach to determining
the characteristics of emitting plasma has serious limitations
(Sylwester, 1990) and a more advanced approach relies on the concept
of differential emission measure, DEM (Sylwester, Schrijver, and
Mewe, 1980). We used the DEM approach to analyze of quiet-Sun RESIK
spectra arranged by activity grouping. The algorithm we used follows
the Bayesian approach, described in detail by Sylwester, Schrijver,
and Mewe (1980), and called the Withbroe-Sylwester algorithm. To
apply this algorithm to the analysis of RESIK data, we selected 15
spectral intervals in all four RESIK channels. The wavelength ranges
of channels 1 and 2 were divided into 2 broad ranges (between
0.15~\AA\ and 0.25~\AA\ wide) with the remaining 11 ranges in
channels 3 and 4 placed around stronger lines or line groups. The
ranges are given in Table~1, together with principal contributors to
the emission. We performed  tests by slightly varying the band
widths to find which were the most suitable.

The spectral fluxes were then integrated over each band to improve
the count statistics. These fluxes were then used as input to the
DEM iterative algorithm. Theoretical spectra providing another
source of input for the DEM inversions have been calculated from
the CHIANTI v5.2 atomic code ({\it SolarSoft}) using the coronal,
photospheric, and the other suitable plasma composition models.
%%%%%%%%%%%%%% begin Fig. 5  ONE COLUMN  %%%%%%%%%%%%%%%%%%%%%%%%%%%%%%%%%%%%%%%%%
%   \begin{figure*}
   \begin{figure}
   \centering
   \includegraphics[width=9 cm]{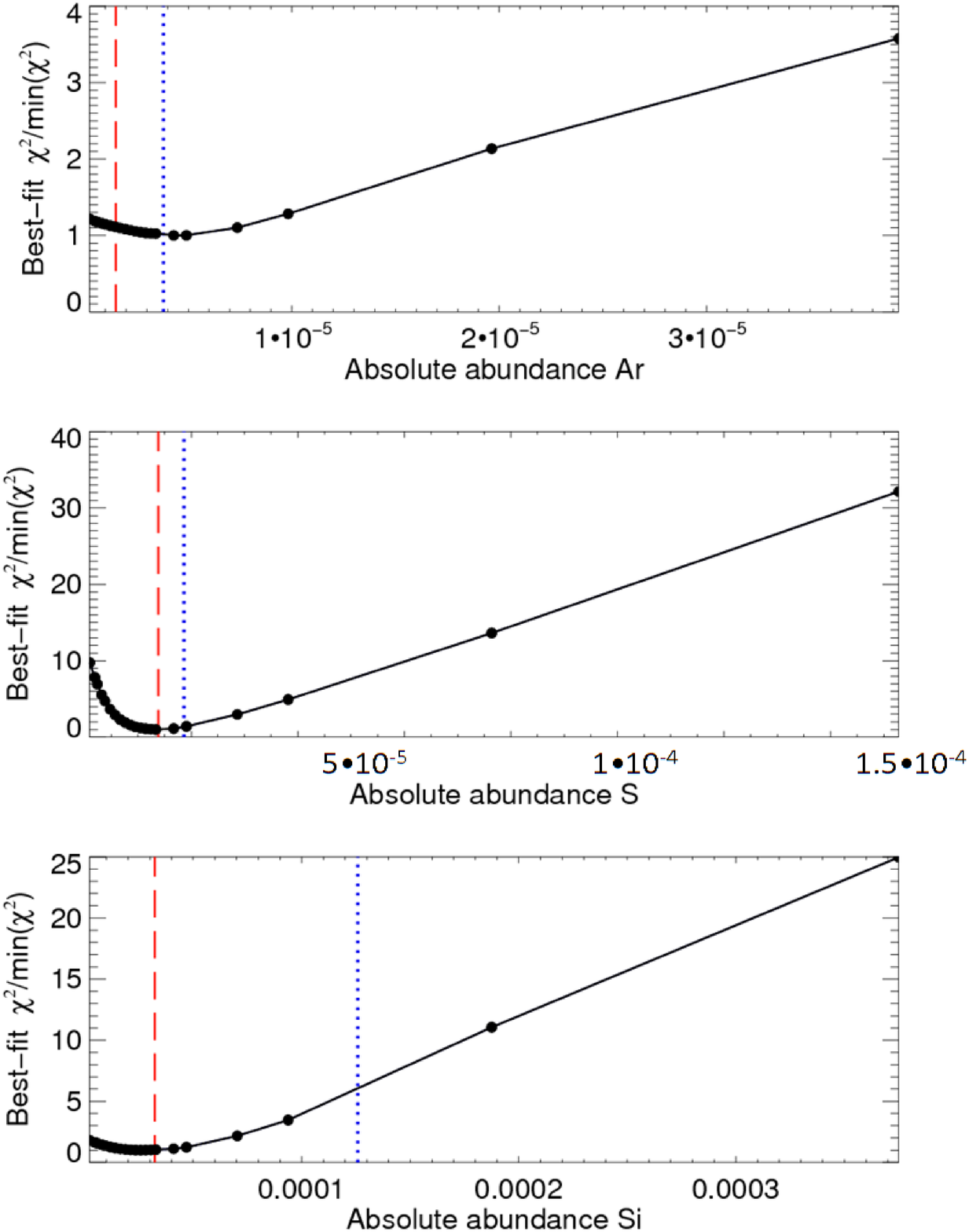}
   \vspace{-1mm}
      \caption{
      Ratios of $\chi^2/min(\chi^2)$ plotted against the absolute elemental abundance for Ar, S, and Si. The vertical blue dotted line is plotted at the value corresponding to the respective coronal abundance of the element, while the red dashed at the photospheric value.
      }
         \label{Figchisq}
%   \end{figure*}
   \end{figure}
%%%%%%%%%%%%%%%%%%%%%%% end Fig. 5 %%%%%%%%%%%%%%%%%%%%%%%%%%%%%%%%

The goodness of the fit between the observed and DEM-fitted band
fluxes was determined by the standard normalized $\chi^2$ values,

\begin{center}
\begin{equation}
\chi^2=\frac{1}{n}\sum_{\rm i=1}^{\rm
n}{\frac{(F_i-F_{ic})^2}{\sigma_i^2}},
\end{equation}
\end{center}

\noindent which is easy to determine when the uncertainties (due to
photon count rate statistics) are known. The absolute fluxes should
have very small uncertainties because the RESIK intensity
calibration is well known (Sylwester et al., 2005). The DEM
iterative process was continued for $10^4$ iterations or whenever
continuous convergence was attained.
% Table 2
\begin{table*}
\caption{Derived plasma parameters for individual activity classes}  % title of Table
\label{table:2}      % is used to refer this table in the text
\centering                          % used for centering table
\begin{tabular}{c c c c c c}        % centered columns (4 columns)
\hline\hline                 % inserts double horizontal lines
\vspace{1mm}
 Level               & {\bf A9-B1} & {\bf B1-B2} & {\bf B2-B3} & {\bf B3-B4} & {\bf B4-B5}   \\ % table
\hline
{\bf Isothermal}         &{\bf results}&             &             &             &           \\
&&&&&\\
\hline
%\vspace{3mm}
&&&&&\\
$T_{\rm D/w3}$ [MK]     & 3.11        &  3.68       & 3.84        & 3.88        & 3.85           \\
\hline
&&&&&\\
$T_{\rm G}$ [MK]      & 7.7         &  6.5        & 5.7         & 5.1         & 5.1           \\
$EM_{\rm G}^a$        & 0.98        &  2.7        & 4.8         & 9.8         & 12.6          \\
$ThM_{\rm G}^b$       & 1.00        &  1.40       & 1.67        & 2.08        & 2.38          \\
\hline
%\vspace{3mm}
&&&&&\\
$T_{\rm 3/4}$ [MK]       & 4.8         &  4.8        & 4.0         & 3.6         & 3.8           \\
$EM_{\rm 3/4}^a$         & 1.6         &  7.5        & 19.7        & 47.8        & 59.7          \\
$ThM_{\rm 3/4}^b$        & 0.80        &  1.70       & 2.33        & 3.29        & 3.81          \\
%\vspace{3mm}
\hline\hline
{\bf Multithermal}           & {\bf results}&             &      &             &               \\
&&&&&\\
\hline
 {\bf El. abun}          & $10^{-6}$   $\Downarrow$  &   &   &   &     \\    % table  heading
Ar                       &  4.65       &   4.59      &  4.35       &  4.45       &  4.27 \\
S                        &  5.62       &  10.6       & 14.3        & 17.6        & 17.8          \\
Si                       & 27.4        &  25.5       & 26.5        & 25.4        & 27.7            \\
%\vspace{1mm}
\hline                 % inserts double horizontal lines
%   {\bf Char.} &    {\bf of}   & {\bf low-T} &{\bf comp.}&               &                 \\
$T_{\rm L}$ [MK]         & 2.9           &  2.6          & 2.5           & 2.4           & 2.3            \\
$EM_{\rm L}^a$  & 0.17           &  0.79          & 2.77           & 8.63           & 22.2 \\
$ThM_{\rm L}^b$  & 1.58           &  3.08          & 5.54           & 9.25           & 14.1 \\           %\vspace{1mm}
\hline                 % inserts double horizontal lines
%   {\bf Char.} &    {\bf of}   & {\bf high-T} &{\bf comp.}&               &                 \\
$T_{\rm H}$ [MK]         & 9.1           &  8.0          & 6.9           & 6.5           & 6.1            \\
$EM_{\rm H}^a$  & 0.023           &  0.089        & 0.42         & 1.03           & 1.86  \\
$ThM_{\rm H}^b$  & 0.18           &  0.32          & 0.58           & 0.87           & 1.11 \\
\hline                 % inserts  horizontal lines
\end{tabular}
{\small \flushleft $^a$ in $10^{47}$~cm$^{-3}$;  $^b$ in
$10^{15}$~g~cm$^{-3/2}$.}
\end{table*}
In Eq.~3, $F_i$ is the photon flux measured in band \emph{i},
$F_{ic}$ is the flux calculated based on given DEM model, \emph{n}
the number of spectral bands used (\emph{n=15}), and $\sigma_i$ the
uncertainty in the measured flux of \emph{i} band. We have initially
performed the DEM inversion over the temperature range from 2~MK to
30~MK, but we found it unnecessary to extend the range beyond
$T>15$~MK. No further improvement in the fit quality is observed
with a larger band-width in the case of the quiet spectra analysis.
Clearly the line and continuum fluxes depend on the plasma
composition of the source region, so an appropriate element
abundance model must be used to recover the DEM shape accordingly.

It is not obvious which of the two generally used composition models
(coronal or photospheric abundances) should be used in the analysis
of the quiet corona spectra, so we did the DEM inversions for both
sets of plasma abundances. The resulting best-fit DEM shapes for the
averaged spectrum (integration time 26.2 hours) are shown in the
left and middle panels of Fig.~4. The envelope of uncertainties of
the DEM shape, from 100 Monte-Carlo runs, is shown by the lengths of
the error bars. It can be seen how sensitive the resulting shape of
the average DEM is to the assumed model of the plasma composition.
The best-fit $\chi^2$ values are beyond an acceptable range for both
photospheric or coronal composition models. We thus concluded that
to provide a reliable estimates of the DEM shape we should also
finely adjust the plasma composition in order to optimize the fit
between the observed and calculated spectral band fluxes. To do
this, we generated a large spectral look-up, many-dimensional
spectral database containing the synthesized profiles in the
spectral ranges covered by RESIK (each 0.001~\AA), in 101
temperature points (equidistant between $6<{\rm log}\,T<8$), with
varying abundances of the elements Ar, S, and Si. The abundances of
the other elements not directly influencing the spectral shape for
the considered $T$-range of calculations were kept equal to their
coronal values. The range of particular element variability was
taken from 0.1  to 20 times the coronal abundance value.

With such a large spectral look-up cube it has been relatively easy
to look for effects of the  dependence of the DEM inversion on
particular-element composition. The results obtained are illustrated
in Fig.~5, where we show the trend in the minimum value of $\chi^2$
as a function of element absolute abundance for Ar, S, and Si. A
well-defined minimum is present in each of the plots. The location
of the minimum along the abundance axis is placed close to the
photospheric value for S and Si. For the element Ar, which has a
high value of first ionization potential (FIP), derived optimum
abundance is above the coronal value. It follows that neither the
photospheric nor the coronal abundance models can be used to
describe the observed set of spectral intensities at the same time.
Therefore we performed the DEM inversion using the optimum abundance
values as determined from the position of the minima in Fig.~5. The
result is shown in Fig.~4 (right panel).  It is seen that the
calculated shape of the DEM is smoother in comparison with the
photospheric or coronal cases and the error envelope decreases. The
calculated value of the normalized $\chi^2$ is also brought now into
an acceptable range. As a consequence of this exercise,  we decided
to perform the composition optimization process for every DEM
inversion for corresponding activity class. The results are shown in
Table~2 and illustrated in Fig.~7.

The calculated, abundance-optimized shapes of DEM distribution are
presented in different colors. The black histogram represents the
highest activity class (B4-B5). No error bars are shown in order to
increase the visibility.  It is observed that all DEM distributions
are bimodal with the cooler (lower temperature) and hotter (higher
temperature) components well separated. As the activity level
increases, the average temperatures of the low and high-T components
shift toward lower temperatures. This is somewhat unexpected but can
be understood as partly due to the deconvolution process working
more effectively when the emission measure ratio of the lower and
higher-T components is decreasing. The amount of cooler plasma found
is orders of magnitude higher than the hotter one. Depending on the
activity level, this amounts to 750 and 1190 for the lower and
higher activity levels, respectively. In Table~2 the basic
characteristics of the cooler (index L) and hotter (index H) plasma
components is presented. In calculating the total thermal energy
content from the obtained DEM=$\varphi(T)$ distributions, we used
the following formulas, derived assuming a constant pressure or
density in the emission volume:
 \begin{equation}
  E_{th}|_{p={\rm const}}=\sqrt{V}\,\, 3 k \sqrt {\int{T^2\varphi(T)dT}}
 \end{equation}
 \begin{equation}
  E_{th}|_{Ne={\rm const}}=\sqrt{V}\,\, 3 k \frac{\int{T \varphi(T) dT}}
  {\sqrt{\int{\varphi(T)dT}}}~~~.
 \end{equation}

\noindent It is straightforward to assign the meaning of
thermodynamic measure $ThM$ to the righthand factors of Eqs.~4 and
5. It was found that the difference between results obtained in the
constant pressure or constant density assumptions did not differ by
more than a few percent for the considered DEM shapes, so in Table~2
we insert average values, characteristic of  the cooler and hotter
components. With the knowledge of respective volumes occupied by a
cooler and the hotter plasma, it would be possible to estimate
thermal energy content of respective components. This study is in
progress.
%%%%%%%%%%%%%% begin Fig. 6  ONE COLUMN  %%%%%%%%%%%%%%%%%%%%%%%%%%%%%%%%%%%%%%%%%
%   \begin{figure*}
   \begin{figure*}[t]
   \centering
   \vspace{5mm}
   \includegraphics[width=15 cm]{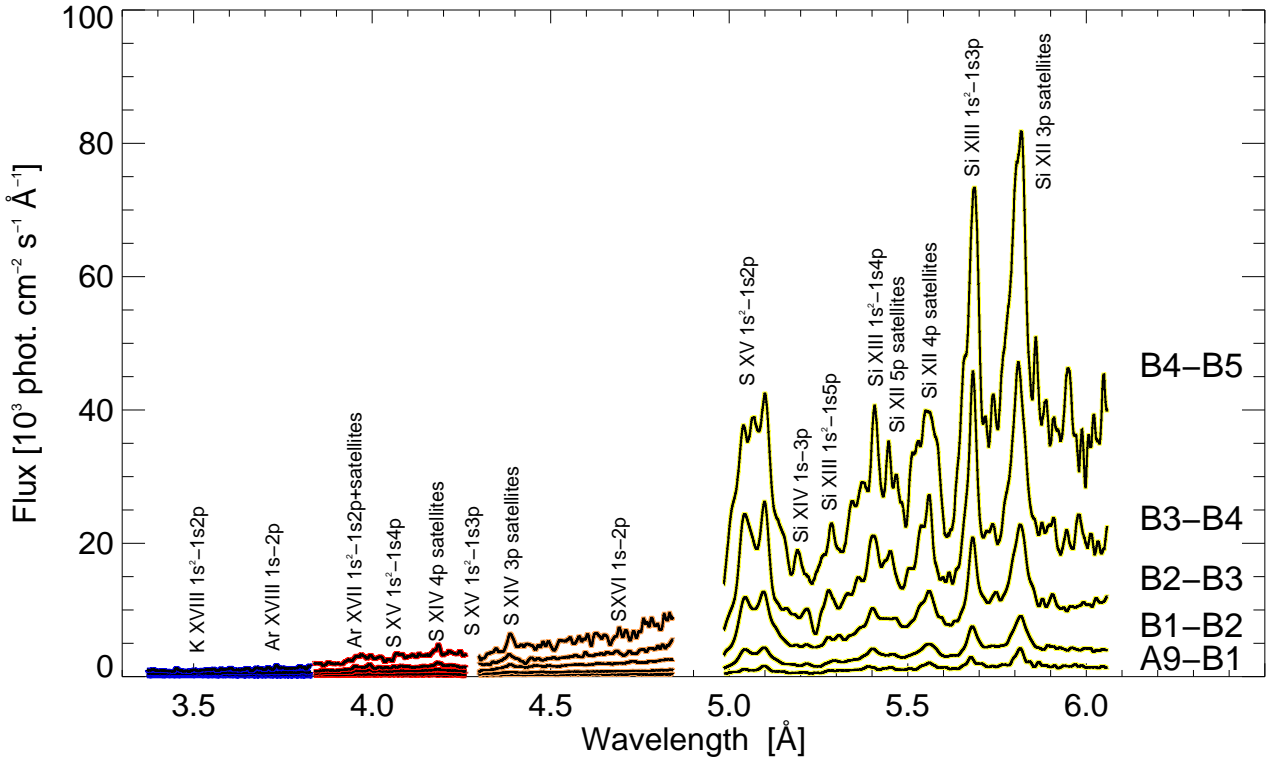}
   \vspace{15mm}
      \caption{The absolute RESIK spectra in linear scale grouped according to
       the \emph{GOES} class. The most prominent lines are indicated.}
         \label{FigVibStab}
%   \end{figure*}
   \end{figure*}
%%%%%%%%%%%%%%%%%%%%%%% end Fig. 6 %%%%%%%%%%%%%%%%%%%%%%%%%%%%%%%%

The results of DEM calculations with the element-abundance
optimization for the five activity classes are shown in Fig.~7 in
different colors. The bimodal character of resulting DEM is evident.
The average values of the  parameters characterizing each of the two
plasma components are given in Table~2 along with the associated set
of optimum elemental abundances is provided. The data set in Table~2
shows that the abundances of Ar and Si do not depend on the activity
level, but the sulfur abundance of the plasma contributing to RESIK
channel 4 spectra systematically increases by a factor of $\sim 4$
from a level well below the photospheric to the coronal abundance
with activity increasing from lower to higher classes. The reason
for this behavior is not known at present. It is important to note
that the optimum abundances of Si are below the generally assumed
photospheric level, but the Ar abundances (Ar is a high-FIP element)
lie even above coronal values. As concerns the energy content
associated with the two plasma components, it is interesting to note
that the thermodynamic measure ratios of the lower and higher-$T$
components are much less than the ratio of respective emission
measures. Provided that the emitting volumes are comparable, this
would mean that the ratio of the total thermal energy content of the
hotter and cooler components is not all that different, and as a
consequence, the heating processes responsible for formation of the
hotter component should not be disregarded in studies of the overall
energy balance even for nonflaring coronal conditions.

\section{Concluding remarks}

%%%%%%%%%%%%%% begin Fig. 7  ONE COLUMN  %%%%%%%%%%%%%%%%%%%%%%%%%%%%%%%%%%%%%%%%%
   \begin{figure}
   \centering
   \includegraphics[width=8 cm]{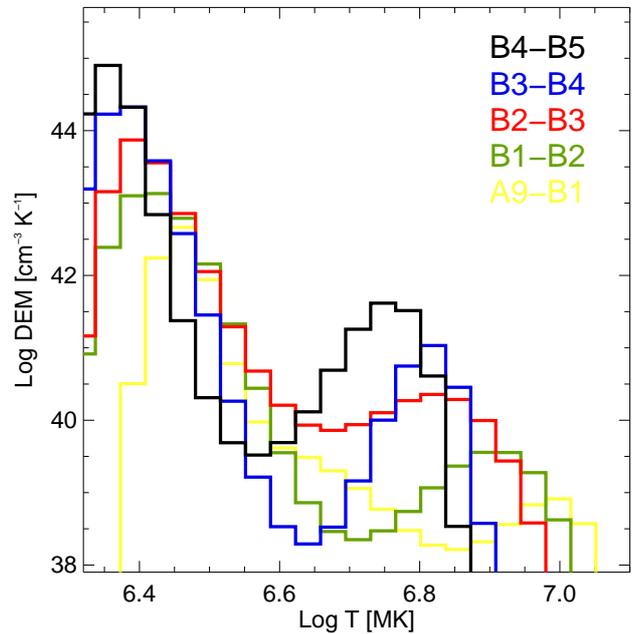}
     \vspace{2mm}
      \caption{
      The histogram representation of differential emission measure (DEM) distributions as obtained from RESIK
      spectra for individual classes of solar activity. Different colors correspond to the following classes:
       black (dark) represents the B4-B5 class and yellow (lightest) A9-B1 class.
      }
         \label{FigVibStab}
   \end{figure}
%%%%%%%%%%%%%%%%%%%%%%% end Fig. 7 %%%%%%%%%%%%%%%%%%%%%%%%%%%%%%%%

We analyzed 312 individual RESIK spectra {grouped} into 5 different
activity classes based on \emph{GOES} levels of activity recorded at
the time of the spectra collection. {All selected spectra were taken
during the nonflaring, low activity conditions prevailing in the
corona, though some weak active-region emission was always present.
The spectral observations (all with a five-minute integration time)
cover the period between 2003 January~1 and March~14.} The analysis
assumed both isothermal and multithermal distributions, the latter
leading to determination of DEM distributions for each activity
class. The DEMs were obtained with the Withbroe-Sylwester Bayesian
iterative, maximum likelihood procedure. The fluxes integrated over
15 wavelength bands covering the range 3.3~-~6.1~\AA\ were used as
an input data for the deconvolution. We found it necessary to allow
for the emitting plasma composition to be nonstandard, i.e. neither
coronal nor photospheric, by varying the abundances of those
elements making important contributions to the line emission in the
spectra. This analysis has never been done before except for a few
flares (to be published, Montreal COSPAR). The main results of the
present study follow.
 \begin{enumerate}
      \item It is necessary to use the multithermal approach in analyzing the spectra. The
      results obtained in the isothermal approximation provide different values of $T$ and $EM$
      parameters depending on the ratio considered. However, the values of so-called thermodynamic
       measure that is directly related to the total thermal energy plasma content are very close,
       because they are surprisingly independent of the particular ratio used for its determination.
      \item It appears necessary to allow for the plasma composition differences when using the
       multitemperature approach in the analysis. Neither coronal nor photospheric composition
        models are able to describe the observed spectra satisfactorily. This result may
         also bias the outcome of any isothermal analysis performed with an isothermal approach.
      \item The two-temperature character of the DEM shape determined here for non-flaring
       plasmas has also been obtained for flares (Sylwester et al. 2008). The presence of the
        higher-$T$ component, with $T$ somewhat below 10~MK, is physically important.
        The emission measure associated with this hotter plasma is $\sim 3$~orders of magnitude
         smaller than in the generally accepted $T \sim 2 - 3$~MK component. This higher-$T$
         component is required because it is impossible to reproduce the observed spectra without it.
   \end{enumerate}

 It is worth noting that the presence of a hotter component in active region emission has been recently
  suggested by Reale et al. (2009) from their analysis of \emph{Hinode} XRT images of an active region and by
   Schmelz et al. (2009), although the very high temperature of the hotter component discussed by Schmelz et al. (2009) has
    been reduced to a value near what is obtained in this work (Schmelz 2009, priv. comm.).
%%%%%%%%%%%%%%%%%%%%end of concluding remarks%%%%%%%%%%%%%%%%%%%%%%%%%%%%

\begin{acknowledgements}
RESIK is a common project between the NRL (USA), MSSL and RAL (UK),
IZMIRAN (Russia), and SRC (Poland). This work was partially
supported by the International Space Science Institute in the
framework of an international working team (No. 108). We acknowledge
the travel support from a U.K. Royal Society/Polish Academy of
Sciences International Joint Project. CHIANTI is a collaborative
project involving the NRL (USA), RAL (UK), MSSL (UK), the
Universities of Florence (Italy) and Cambridge (UK), and George
Mason University (USA). The research leading to these results
received partial funding from the European Commission's Seventh
Framework Programme (FP7/2007-2013) under grant agreement No. 218816
(SOTERIA project, www.soteria-space.eu).

\end{acknowledgements}


\begin{thebibliography}{}

\bibitem[1996]{bros96} Brosius, J.W., Davila, J.M., \& Thomas, R.J. 1996, ApJS, 106, 143
\bibitem[1998]{bros98} Brosius, J.W., Davila, J.M., \& Thomas, R.J. 1998, ApJS, 119, 255

\bibitem[2007]{chif07} Chifor, C., del Zanna, G., Mason, H.E., Sylwester, J., Sylwester, B. \& Phillips, K. J.H. 2007, A\&A, 462, 323
\bibitem[2004]{curd04} Curdt, W., Landi, E. , \& Feldman, U. 2004, A\&A, 427, 1045
\bibitem[1995]{dela95} Delaboundiniere, J.-P., Artzner, G.E., Brunaud, J.,
        et al. 1995, Sol. Phys., 162, 291
\bibitem[1971]{dup71} Dupree, A. K., \& Reeves, E. M. 1971, ApJ, 165, 599
\bibitem[1973]{dup73} Dupree, A. K., Huber, M.C.E., Noyes, R.W., Parkinson, W.H.,
        Reeves, E. M., \&  Withbroe, G.L. 1973, ApJ, 182, 321
\bibitem[1998]{free98} Freeland, S.L., \& Hardy, B.N. 1998, Sol. Phys., 182, 497
\bibitem[2007]{han07} Hannah, I.G., Hurford, G.J., Hudson,
        H.S., Lin, R.P., \& van Bibber, K. 2007, ApJ, 659, L77
\bibitem[2007]{klim08} Klimchuk, J.A., Patsourakos, S., \& Cargill, P.J. 2008, ApJ, 682,
1351
\bibitem[2004]{kretz04} Kretzschmar, M., Lilensten, J., \& Aboudarham,
        J. 2004, A\&A, 419, 345

\bibitem[1973]{mali73} Malinovsky, M., \& Heroux, L. 1973, ApJ, 181, 1009
\bibitem[1966]{pet66} Peterson, L.E., Schwartz, D.A., Pelling, R.M.,
        \& Mckenzie, D. 1966, J.Geophys. Res., 71, 5778

\bibitem[2006]{phil06} Phillips, K.J.H., Dubau, J., Sylwester, J. \& Sylwester, B. 2006,
            ApJ, 638, 1154
\bibitem[2007]{ral07} Ralchenko, Yu., Feldman,U. \& Doschek, G.A. 2007, ApJ, 659, 1682
\bibitem[2009]{reale09} Reale, F., Testa, P., Klimchuk, J.A. \& Parenti, S. 2009, ApJ, 698, 756
\bibitem[2009]{schmelz09} Schmelz, J.T., Saar, S.H., DeLuca, E.E., Golub, L., Kashyap, V.L., Weber, M.A., \& Klimchuk, J.A.
  2009, ApJ, 693, L131
\bibitem[2000]{Shea00} O'Shea, E., Gallagher, P.T., Mathioudakis, M., Phillips, K.J.H., Keenan F.P. \& Katsiyannis 2000,
A\&A, 358, 741
\bibitem[1980]{Syl80} Sylwester, J., Schrijver, J. \& Mewe, R. 1980, Sol. Phys., 67, 285
\bibitem[1990]{Syl90} Sylwester, J. 1990, The Dynamic Sun, Proceedings of the 6th European Meeting on Solar Physics, Debrecen, 21-24 May, 1990. Edited by L. Dezso.  Publications of Debrecen Heliophysical Observatory of the Hungarian Academy of Sciences, Vol. 7, 212
\bibitem[2005]{Syl5} Sylwester, J., Gaicki, I., Kordylewski, Z., \& 19 others 2005,
            Sol. Phys., 226, 45
\bibitem[2006]{syl6} Sylwester, B., Sylwester, J., Siarkowski, M., Phillips, K.J.H.,  Culhane,J.L., Lang,J., Brown,C. \& Kuznietsov,V.D. 2006, Adv. Space Res., 38, 1534
\bibitem[2006]{syl6a} Sylwester, B., Sylwester, J., Kepa, A., Kordylewski, Z., Phillips, K.J.H.,  \& Kuznietsov,V.D. 2006, Sol. System Res., 40, 125
\bibitem[2008]{syl8} Sylwester, B., Sylwester, J. \& Phillips, K.J.H. 2008, J. Astrophys. Astr., 29, 147
\bibitem[2008]{syl9} Sylwester, J., Sylwester, B. \&
Phillips, K.J.H. 2008, ApJ, 681, L117
\bibitem[1998]{warren98} Warren, H.P.,  Mariska,J.T. \& Lean, J.,
         1998, J. Geoph. Res., 103, 12077 and 12091
\bibitem[1978]{ver78} Vernazza, J.E. \& Reeves, E.M. 1978, ApJS, 37, 485
\bibitem[2007]{you07} Young, P.R., Del Zanna, G., Mason,
H.E., \&  9 others 2007, PASJ, 59, S857





\end{thebibliography}
\end{document}